\documentclass{article}

\usepackage[T1]{fontenc}
\usepackage[utf8]{inputenc}
\usepackage{arxiv}
\usepackage{booktabs}
\usepackage{amsfonts}
\usepackage{microtype}
\usepackage{graphicx}
\usepackage[numbers,square,sort&compress]{natbib}
\usepackage{array}
\usepackage{amsmath, amssymb}
\usepackage{float}
\usepackage{algorithm}
\usepackage{algpseudocode}
\usepackage{xurl}
\usepackage{hyperref}
\usepackage{doi}
\renewcommand{\doi}[1]{\begingroup\urlstyle{rm}\href{https://doi.org/#1}{doi:\nolinkurl{#1}}\endgroup}

\algrenewcommand{\algorithmicrequire}{\textbf{Input:}}
\algrenewcommand{\algorithmicensure}{\textbf{Output:}}

\title{FSSDataBase: a reconstructable dataset of simulated frequency-selective surface structures and scattering responses}
\renewcommand{\shorttitle}{FSSDataBase: a reconstructable dataset}

\author{
  Xinke Kuang \\
  Hangzhou Institute of Technology, Xidian University\\
  HangZhou, China \\
  \texttt{25241215209@stu.xidian.edu.cn} \\
  \And
  Shiyun Ma \\
  Hangzhou Institute of Technology, Xidian University\\
  HangZhou, China \\
  \texttt{25241214964@stu.xidian.edu.cn} \\
\And
  Yuanyuan Wang \\
  Hangzhou Institute of Technology, Xidian University\\
  HangZhou, China \\
  \texttt{25241215318@stu.xidian.edu.cn} \\
\And
  Jiang Wu \\
  Hangzhou Institute of Technology, Xidian University\\
  HangZhou, China \\
  \texttt{25241215366@stu.xidian.edu.cn} \\
}

\begin{document}
\maketitle

\begin{abstract}
Data-driven design of frequency-selective surfaces (FSSs) requires reusable datasets that link structural geometry to electromagnetic response under documented simulation conditions.
Here we present FSSDataBase, an openly available collection of 5,000 procedurally generated single- and multilayer FSS unit cells simulated using Ansys HFSS.
The dataset covers 10-20 GHz with transverse-electric and transverse-magnetic relfection and transmission responses, including magnitude and phase, for recorded incidence angles of $0^{\circ}$ and $30^{\circ}$.
Each record links binary structural masks and JSON-based reconstruction metadata to raw response samples and derived magnitude labels.
Structural descriptors, response-distance measures and task-oriented cost functions characterize geometric and electromagnetic variation and support screening for polarization stability, angular-selective bandwidth and matched-amplitude, $180^{\circ}$ phase-separated structure pairs.
Accompanying code supports model reconstruction, data processing and configuration-driven generation of additional samples.
Beyond conditional forward modelling and response-based structure retrieval, FSSDataBase has the potential to serve as a reference dataset for deep-learning-based FSS inverse design, supporting fair comparisons across network architectures under shared data splits,input conditions and evaluation protocols.
\end{abstract}


\section{Background \& Summary}
\label{sec:background-&-summary}
Frequency-selective surfaces (FSSs) are periodic or quasi-periodic arrangements of conducting and dielectric elements whose spatial filtering response arises from element resonances, lattice interactions, and multilayer propagation effects \cite{ref01}.
Their reflection and transmission behavior is determined jointly by element geometry, periodicity, substrate and superstrate stack, operating frequency, incidence angle, and polarization \cite{ref02}.
A scattering response may be represented as $R=F\left( G,f,M,\theta ,p,\mathcal{B}\right)$, where $G$ denotes geometry, $f$ frequency, M material and layer configuration, $\theta $ incidence angle, $p$ polarization, and $\mathcal{B}$ the boundary-condition and excitation specification.
Mesh and solver settings should additionally be retained to make a converged numerical result reproducible.
Consequently, a single geometry may yield multiple valid response curves when evaluated under different materials, angles, polarizations, or boundary conditions.
Exploring this couple space with high-fidelity full-wave simulations commonly requires repeated solves over frequency and excitation conditions \cite{ref02}.

This computational burden has motivated data-driven models.
Representation-learning frameworks have been used for FSS inverse design \cite{ref03}, while combinations of equivalent-circuit models and deep learning have been proposed for wideband FSS design \cite{ref04}.
Inverse design, however, has a different form of non-uniqueness: multiple geometries may satisfy a similar target response.
This should not be conflated with the forward conditional relationship above.
Forward modelling maps a specified geometry-condition tuple to its response, whereas inverse design searches a geometry space for structures consistent with a target under stated conditions \cite{ref03}.
Distinguishing these relationships is important for reproducible model development and for evaluating whether a predicted geometry physically valid under its intended operating conditions.

Published FSS studies illustrate the breadth of geometry- and condition-dependent behavior.
Highly selective FSSs with ultrawideband rejection \cite{ref05}, ultra-wide-angle bandpass responses \cite{ref06}, angularly stable tunable designs \cite{ref07}, polarization-rotating structures \cite{ref08}, and dual-polarized angle-selective double-layer surfaces \cite{ref09} each address different portions of the design and excitation space.
Miniaturized structures \cite{ref10}, three-dimensional bandpass configurations \cite{ref11}, tunable absorptive/transmissive surfaces \cite{ref12}, and angularly and polarization-stable FSSs \cite{ref13} further show that compactness, bandwidth, reconfigurability, and robustness are generally stack- and geometry-specific.
Results reported under one material, excitation, or boundary specification therefore cannot automatically be transferred to another, which makes complete condition metadata essential for reuse.

Condition sensitivity also arises in frequency-reconfigurable multilayer FSSs \cite{ref14}, flexible three-dimensional structures \cite{ref15}, ultrathin polarization rotators \cite{ref16}, actively controlled dual-band surfaces \cite{ref17}, and frequency-dependent coupling designs \cite{ref19}.
These examples demonstrate that material state, layer configuration, deformation, active bias, and electromagnetic coupling contribute to the response alongside static geometry.
Recent reviews of neural-network methods for photonic and electromagnetic devices likewise show that modelling requirements grow when material dispersion, incidence and polarization variation, and data-processing choices must be represented explicitly \cite{ref19,ref20}.

Learning-based inverse methods continue to develop through diffusion models \cite{ref21}, transfer-learning strategies \cite{ref22}, and integrated forward-inverse AI workflows \cite{ref23}.
Resources assembled for such modelling studies can nevertheless be difficult to reuse when geometric encoding are task-specific; material, solver, boundary-condition.
Incidence-angle, and polarization metadata are incomplete or heterogeneous; reconstruction information such as coordinate conventions and layer ordering is unavailable; multi-condition responses for a common geometry are limited; or data splits and validation protocols are not comparable.

Retaining explicit geometry-condition-response associations addresses these limitations.
It enables reproducible benchmarking under identical descriptors and split definitions; allows conditional surrogate models to learn compatible response mappings; help inverse-design workflows distinguish physically different solutions; and supports tests of robustness and cross-condition generalization across angle, polarization, material configuration, and other defined conditions.
Such documentation does not itself guarantee superior performance or coverage, but it makes the basis for subsequent reuse inspectable and reproducible.

FSSDataBase is designed as a 5,000-sample collection of simulated FSS structures and their scatter responses.
The dataset covers the 10-20GHz range and records responses under transverse-electric (TE) and transverse-magnetic (TM) excitations at incidence angles if $0^{\circ}$ and $30^{\circ}$.
Each dataset recording links structural information to raw scattering data, enabling response curves to be interpreted together with their associated geometry and operating conditions rather than as isolated labels.

To support reconstruction and reuse, each record includes a binary structural representation, raw response files, derived response labels, and a JSON-based description that preserves the information required to reconstruct the simulated structure.
The dataset and associated code are openly available.
This organization supports conditional forward modelling, inverse-design studies, angular and polarization robustness analysis, representation-learning benchmarks, and reproducible comparisons among data-driven electromagnetic-design workflows.
It also allows users to select subsets defined by structure, response, or excitation conditions while retaining the provenance needed to interpret whose subsets correctly.

\section{Method}
\label{sec:method}
\subsection{Simulation Configuration}
\label{subsec:simulation-configuration}
Electromagnetic simulations were performed in Ansys HFSS 2023 R1 using Floquet-port excitation and master/slave periodic boundary conditions.
Excitation was evaluated at incidence angles of $0^{\circ}$ and $30^{\circ}$.
A frequency sweep from 10 to 20GHz was configured with 101 frequency points.
The F4BM348 substrate was assigned $\varepsilon _{r}=3.48$, a loss tangent of 0.0037, and a thickness of 1mm, and the unit-cell lateral dimension was set to 10mm.
The adaptive solution was limited to a maximum of 12 adaptive iterations/passes with Max Delta $S=0.03$.
All other mesh-generation and convergence settings were left at the HFSS defaults.

\subsection{Dataset-generation workflow}
\label{subsec:dataset-generation-workflow}
\begin{figure}[htbp]
\centering
\includegraphics[width=\linewidth]{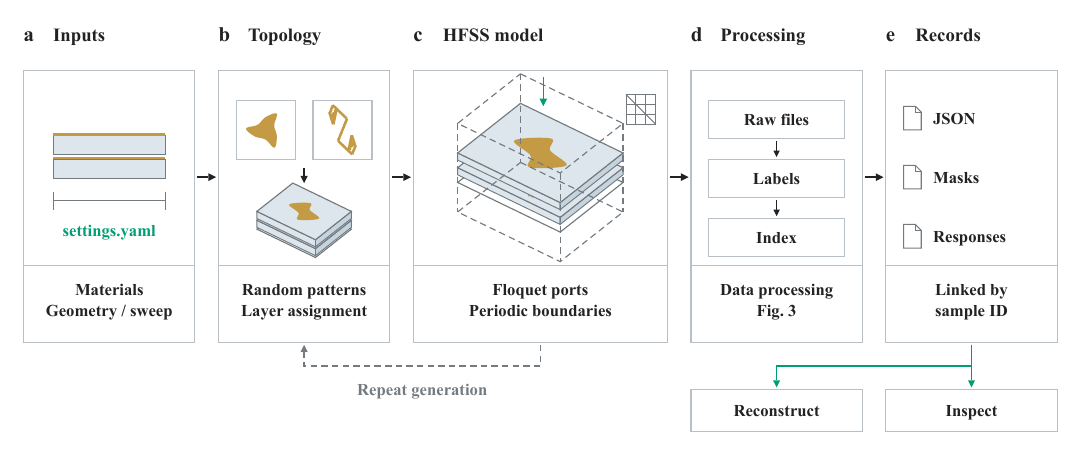}
\caption{Workflow of dataset generation. Configurable material, geometric and excitation parameters define the design space (a).
Metal patterns and layer arrangements are generated procedurally (b), assembled into a periodic HFSS unit-cell model and solved in an independent simulation session for each attempt (c).
Raw results and derived representations are organized through the processing workflow (d), and linked by a common sample identifier in the resulting dataset (e).
The dataset supports subsequent model reconstruction and structure-response inspection.
The multilayer illustration uses the stored layer masks of record 1786866648.2374983; the perspective and air-region extent are schematic.
Metal is illustrated in ochre, dielectric layers in pale blue-grey, and the air region by dashed outlines.
The model uses perfect-electric-conductor metal boundaries.}
\label{fig:1}
\end{figure}

Each sample was generated by stochastically constructing a multilayer FSS stack-up.
As shown in Fig.~\ref{fig:1}, the generation procedure sampled the number and placement of dielectric and metallic layers, dielectric-layer thickness, and internal metallic patterns, while constraining the number of metallic layers to be no greater than the number of dielectric layers plus one.
Internal metallic geometries were constructed from regular-hexagonal subdivisions to enable equal-scale partitioning of the design region.
Samples without a valid metallic geometry were discarded before electromagnetic simulation.

Valid structures were converted into periodic HFSS unit-cell models and simulated using the configuration described in the Simulation Configuration section.
For each accepted structure, the workflow retained continuous TE and RM reflection and transmission responses, including $S_{11}$ and $S_{21}$ magnitudes and phases, together with structural metadata required to reconstruct the model.
The released code and configuration files provide the implementation-level details of model construction, execution control, and error handling.

\subsection{Random topology-generation rules}
\label{subsec:random-topology-generation-rules}

\begin{figure}[htbp]
\centering
\includegraphics[width=\linewidth]{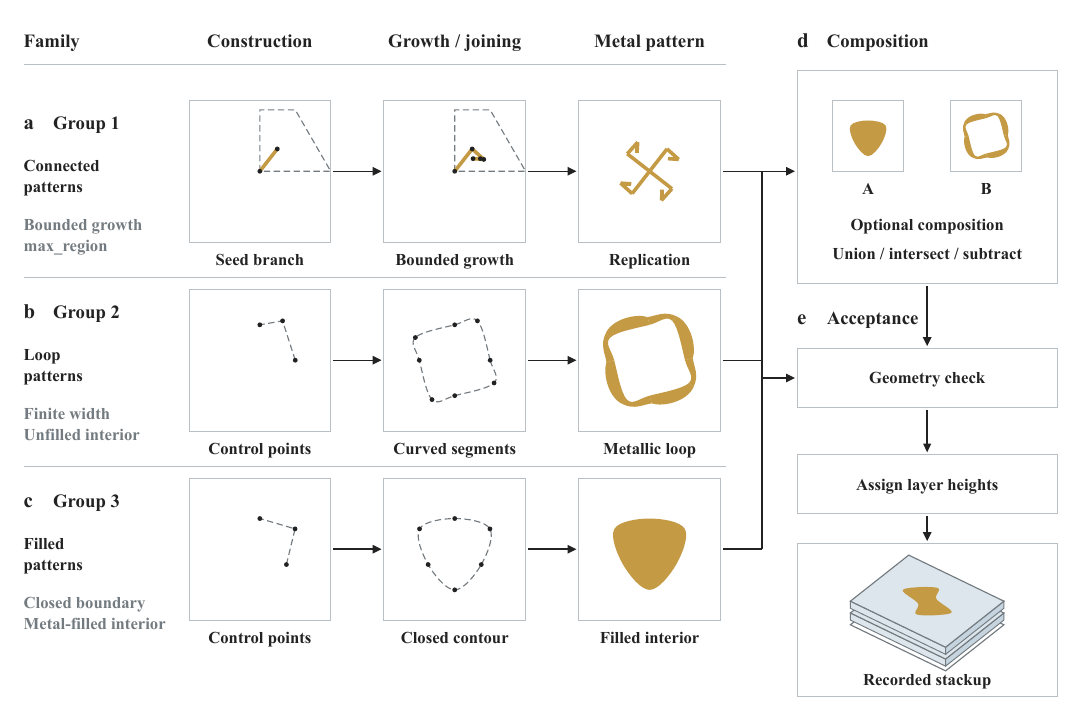}
\caption{Process of topology generation. Three procedural families illustrate controlled geometric randomness: central-connected patterns built from bounded branch growth and rotational replication (a), loop patterns formed from control-point connections (b), and filled patterns constructed from closed contours (c). Black markers denote control points, dashed lines denote construction guides, and ochre denotes generated metal regions. The illustrated connections in (b) are curved; the implementation also supports straight connections. Optional composition combines scaled component patterns using union, intersection or subtraction (d), followed by geometry-validity checking and assignment to metal-layer heights (e). These operators are shown as procedural options, not as evidence of fabrication validation. The standalone generation examples use fixed seeds 20260914--20260916, a 10 mm period and a 0.25 mm linewidth parameter; they have not been newly simulated. The multilayer inset is the separate stored record 1786866648.2374983 and is not claimed to contain the three newly illustrated patterns. Its drawing shares the perspective used in Figs. 1 and 4.}
\label{fig:2}
\end{figure}

As shown in Fig~\Ref{fig:2}, the topology-generation framework comprises three geometric generators: centrally connected branches (Group 1), closed metal loops (Group 2), and filled metal patches (Group 3).
The generators use prescribed sampling regions derived from a regular hexagonal reference geometry to organize the construction of internal metallic patterns.
A rotational repetition order, n, determines how the initial geometry is extended around the pattern central.
When random selection is enabled, n is sampled uniformly from \{2,3,4\}; otherwise, the configured value is used.
The sampling regions define the internal geometry-generation rules, while the periodic simulation cell is specified separately.

\subsubsection{Group 1: Centrally connected branches}

Group 1 constructs a connected metallic pattern by growing a sequence of straight segments from the central.
The initial segment has a length of 0.4a, where a denotes the geometric scale specified by unit.size, and its orientation is randomly selected within the range associated with the chosen repetition order.
Subsequent segments extend from the endpoint of the preceding segment.
Candidate directions are sampled at integer-degree resolution and rejected when their numerical difference from the preceding direction is at most $45^{\circ}$.
Intersection tests against the sampling-region boundary and previously generated segments that do not share the starting point determine the available extension distance.
An extension is rejected when this distance does not exceed 0.1a; otherwise, its length is sampled uniformly within the available distance.

Growth terminates when the estimated total strip area, accounting for rotational replication, exceeds a configured fraction of the reference hexagonal area, or after five consecutive unsuccessful extension attempts.
This area estimate is a growth-control parameter rather than an exact measure of the metal coverage, The accepted segments are converted into rectangular metal strips, replicated at angular intervals of $360^{\circ}/n$, and united to form the centrally connected pattern.
The sequential growth, termination, and rotational replication procedures are summarised in Algorithm~\ref{alg:central-branches}.

\begin{algorithm}[htbp]
\caption{Generation of a centrally connected metal pattern}
\label{alg:central-branches}
\small
\begin{algorithmic}[1]

\Require Geometric scale $a$, trace width $w$,
         area-control fraction $\rho$,
         and branch-number configuration
\Ensure Centrally connected metal pattern $M$

\State $n \gets \operatorname{Select}
       (\{2,3,4\}, \text{branch-number configuration})$
\State $\Omega \gets$ prescribed growth region for $a$ and $n$
\State $A_{\mathrm{hex}} \gets 3\sqrt{3}\,a^2/2$
\State $L \gets \varnothing$, $p \gets (0,0)$,
       $Q \gets 0$, $f \gets 0$

\Repeat
    \If{$L = \varnothing$}
        \State $\theta \gets
        \begin{cases}
            \operatorname{RandomInteger}(0,180), & n=2,\\
            \operatorname{RandomInteger}(0,90),  & \text{otherwise}
        \end{cases}$
        \State $\ell \gets 0.4a$
    \Else
        \State Sample $\theta$ from $\{0,\ldots,359\}$ until
               $|\theta-\theta_{\mathrm{previous}}|>45$
        \State $d \gets$ nearest positive ray-intersection
               distance from $p$
        \Statex \hspace{\algorithmicindent}
                along $\theta$ to $\partial\Omega$ or a previous
                segment not sharing $p$
        \If{no eligible intersection exists or $d \leq 0.1a$}
            \State $\ell \gets 0$
        \Else
            \State $\ell \gets \operatorname{Uniform}(0,1)\times d$
        \EndIf
    \EndIf

    \If{$\ell > 0$}
        \State Append segment $(p,\theta,\ell)$ to $L$
        \State $p \gets p+\ell(\cos\theta,\sin\theta)$
        \State $\theta_{\mathrm{previous}} \gets \theta$
        \State $Q \gets Q+w\ell$
        \State $f \gets 0$
    \Else
        \State $f \gets f+1$
    \EndIf
\Until{$nQ>\rho A_{\mathrm{hex}}$ or $f=5$}

\State $B \gets$ union of rectangular strips of width $w$
       along segments in $L$
\State $M \gets
       \displaystyle\bigcup_{j=0}^{n-1}
       \operatorname{Rotate}\!\left(B,\frac{360^\circ j}{n}\right)$
\State \Return $M$

\end{algorithmic}
\end{algorithm}

\subsubsection{Group 2: Closed metal loops}
Group 2 generates a closed contour from an initial sequence of control points.
In addition to the repetition order, the generator selects a straight-segment or curved representation and one, two, or three intermediate control points when the corresponding randomization options are enabled.
For the two- and three-point cases, points are sampled uniformly within predefined triangular subregions.
The initial point sequence is rotationally replicated, repeated junction points are removed, and the sequence is closed.

For the straight-segment representation, adjacent points are connected using rectangular metal strips of the specified width.
For the curved representation, a closed Catmull-Rom spline is constructed from the control points.
A second curve is obtained through a derivative-based offset controlled by the configured trace parameter.
Corresponding segment endpoints are connected, and the enclosed strip regions are converted into surfaces and united to form the metallic loop.
Because the offset uses unnormalized derivatives, the trace parameter does not impose a strictly constant physical width along the curved contour.
Algorithm ~\ref{alg:closed-loops} summarize control-point sampling, rotational closure, and the construction of straight-segment or curved metallic loops.

\begin{algorithm}[htbp]
\caption{Generation of a closed metal-loop pattern}
\label{alg:closed-loops}
\small
\begin{algorithmic}[1]

\Require Geometric scale $a$, trace parameter $w$,
         and configurations for contour strategy,
         repetition order, and control-point count
\Ensure Closed metal-loop pattern $M$

\State $s \gets \operatorname{Select}
       (\{\text{straight},\text{curved}\},
       \text{contour-strategy configuration})$
\State $n \gets \operatorname{Select}
       (\{2,3,4\}, \text{repetition-order configuration})$
\State $k \gets \operatorname{Select}
       (\{1,2,3\}, \text{control-point-count configuration})$
\State $P \gets$ two prescribed endpoints for $a$ and $n$

\If{$k=1$}
    \State $p \gets$ uniformly sampled point on the
           prescribed radial segment
    \State Insert $p$ immediately after the first point of $P$
\Else
    \ForAll{the $k$ prescribed triangles with vertices $A,B,C$}
        \State $r_1,r_2 \gets$ independent
               $\operatorname{Uniform}(0,1)$ draws
        \State $u \gets \sqrt{r_1}$
        \State $p \gets (1-u)A+u(1-r_2)B+ur_2C$
        \State Insert $p$ immediately after the first point of $P$
    \EndFor
\EndIf

\State $P \gets$ concatenation of rotational copies of $P$
       at angles $360^\circ j/n$
\Statex \hspace{\algorithmicindent}
        for $j=0,\ldots,n-1$, omitting repeated junction points
\State Close $P$ by appending its first point

\If{$s=\text{straight}$}
    \State $M \gets$ union of rectangular strips of width $w$
    \Statex \hspace{\algorithmicindent}
            connecting consecutive points in $P$
\Else
    \State Construct periodic Catmull--Rom segments
    \Statex \hspace{\algorithmicindent}
            $c_j(t)=(x_j(t),y_j(t))$ through $P$,
            with $0\leq t\leq 1$
    \ForAll{segments $c_j$}
        \State $o_j(t) \gets c_j(t)
               +w\bigl(y'_j(t),-x'_j(t)\bigr)$
        \State Connect corresponding endpoints of $c_j$ and $o_j$
        \State $S_j \gets$ surface enclosed by the two curves
               and endpoint links
    \EndFor
    \State $M \gets \displaystyle\bigcup_j S_j$
\EndIf

\State \Return $M$ if surface construction succeeds;
\Statex \hspace{\algorithmicindent}
        otherwise report a construction failure

\end{algorithmic}
\end{algorithm}

\subsubsection{Group 3: Filled metal patches}
Group 3 adopts the control-point sampling and rotational construction rules of Group 2 to define a closed patch boundary.
In the straight-segment variant, the boundary consists of consecutive linear segments; in the curved variant, it follows the Catmull-Rom contour.
The boundary segments are joined, and the enclosed region is covered to create a filled planar metal patch.
Variations in the sampled control points, repetition order, and contour representation therefore produce patches with different outlines while retaining the same construction procedure.
Algorithm~\ref{alg:filled-patches} summarize the generation of filled patches using the shared contour-construction procedure.

\begin{algorithm}[htbp]
\caption{Generation of a filled planar metal patch}
\label{alg:filled-patches}
\small
\begin{algorithmic}[1]

\Require Geometric scale $a$ and Group~3 configurations
         for contour strategy, repetition order,
         and control-point count
\Ensure Filled planar metal patch $M$

\State Obtain $s$ and the closed point sequence $P$
       using the sampling
\Statex \hspace{\algorithmicindent}
        and rotational-closure steps of
        Algorithm~\ref{alg:closed-loops},
\Statex \hspace{\algorithmicindent}
        with the Group~3 configurations

\If{$s=\text{straight}$}
    \State $C \gets$ closed boundary formed by
           consecutive straight segments in $P$
\Else
    \State $C \gets$ closed Catmull--Rom contour through $P$
\EndIf

\State Join the boundary segments of $C$
\State $M \gets$ cover the region enclosed by $C$
\State \Return $M$ if covering succeeds
       and no residual line objects remain;
\Statex \hspace{\algorithmicindent}
        otherwise report a construction failure

\end{algorithmic}
\end{algorithm}

\subsection{Data processing}
\label{subsec:data-processing}
\begin{figure}[H]
\centering
\includegraphics[width=\linewidth]{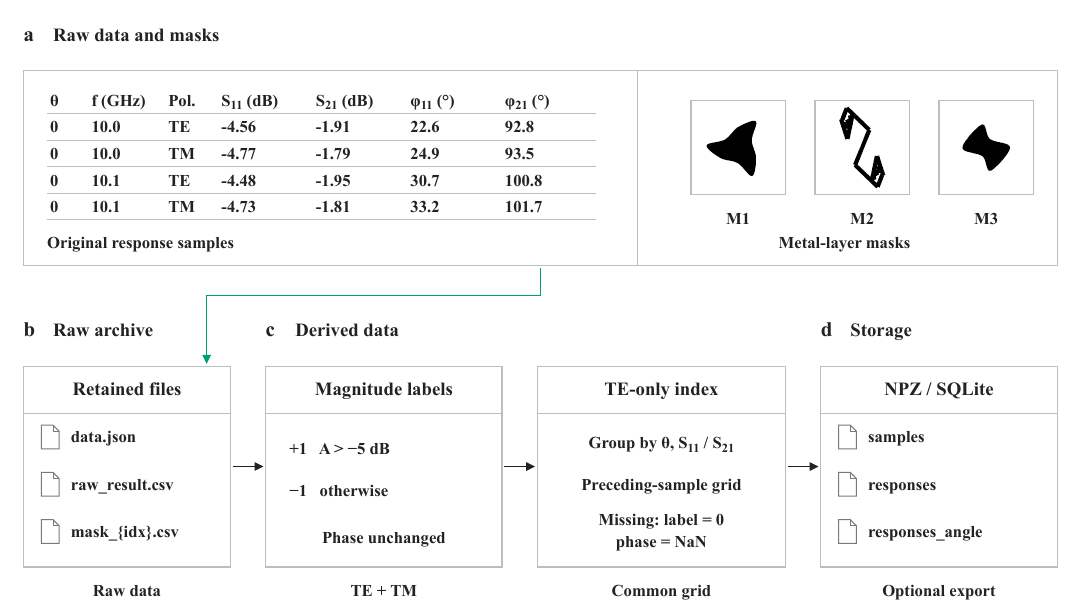}
\caption{Workflow of data processing.
Selected rows of the original response file and the structural metadata illustrate the inputs to data organisation (a).
The raw archive retains the record-level JSON, continuous-valued response samples and binary structural masks (b).
Derived magnitude labels use +1 for values greater than \ensuremath{-}5 dB and \ensuremath{-}1 otherwise, while phase values remain numerical quantities.
Raw and labelled CSV files retain TE and TM channels; the current compact indexing path selects TE responses, groups them by incidence angle and reflection/transmission channel, and uses preceding-sample selection for common-grid alignment (c).
In the compact representation, 0 denotes an unavailable label and NaN denotes unavailable phase. Optional NPZ and SQLite storage retain record identifiers and response-condition associations (d). Black in the displayed masks denotes stored value 1, corresponding to metal; white denotes stored value 0.
This display reverses the source PNG convention of white (255) for metal without changing the underlying mask values. The raw records remain available independently of the derived index. The miniature table is rounded for readability; the source files and Fig. 5 curves retain the exported numerical precision.}
\label{fig:3}
\end{figure}

As illustrated in Fig.~\ref{fig:3}, the exported HFSS responses and structural metadata were organized into a record-level archive and derived representations for indexing and retrieval.
Each record was linked by its sample identifier to the reconstruction metadata, numerical response files, and binary structural masks (Fig.~\ref{fig:3} a,b).
The raw responses retained the original frequency amd incidence-angle samples, together with the $S_{11}$ and $S_{21}$ magnitudes in decibels and phases in degrees for both TE and TM polarizations.
Structural masks were retained separately for each indexed metal pattern, with values of 1 and 0 indicating metal and background, repsectively.

Magnitude labels were derived by applying a common threshold to each response channel:

\begin{equation}
\label{eq:1}
    L_{s,c,p,\theta}(f)=\left\{
\begin{array}{rcl}
+1,       &      & {A_{s,c,p,\theta}(f)      >      -5dB,}\\
-1,     &      & {otherwise,}
\end{array} \right.
\end{equation}

where $A_{s,c,p,\theta}(f)$ denotes the magnitude in decibels for sample $s$, scattering channel $c\in\{S_{11},S_{21}\}$, polarization $p\in\{TE,TM\}$, and incidence angle $\theta$.
Thresholding was performed before numerical values were rounded for CSV export.
The resulting \textit{label\_result.csv} retained the frequency and angle coordinates and both polarization channels, while phase values remained numerical quantities without thresholding (Fig.~\ref{fig:3} c).

The compact indexing procedure selected the TE responses and grouped them by sample identifier, incidence angle, and reflection or transmission channel.
Within each group, responses were ordered by frequency were taken from the source node at or immediately below that frequency.
The current compact representation uses a uniform 300-point indexing grid spanning 1-30 GHz, whereas the simulated responses cover 10-20 GHz.
Unavailable entries were represented by 0 for magnitude labels and NaN for phase in the numerical arrays; missing phase values were stored as NULL in SQLite.
These markers distinguish unavailable entries from the two magnitude-label classes.

\subsection{HFSS reconstruction representation}
\label{subsec:hfss-reconstruction-representation}
To support record-level reconstruction and reuse, the released dataset is accompanied by code that interprets the reconstruction metadata stored in each sample's data.json file.
This information enables users to replay the recorded design history in HFSS and inspect the resulting model definition.

Each reconstruction record separates descriptive metadata from an ordered build-operation log.
The descriptive metadata includes the author, generation timestamp, unique project identifier, material name and electromagnetic properties, substrate material and height, frequency-range setup with the associated solution-point count, and angular range.
The build field records the chronological geometric operations used to define the structure.

\begin{figure}[htbp]
\centering
\includegraphics[width=\linewidth]{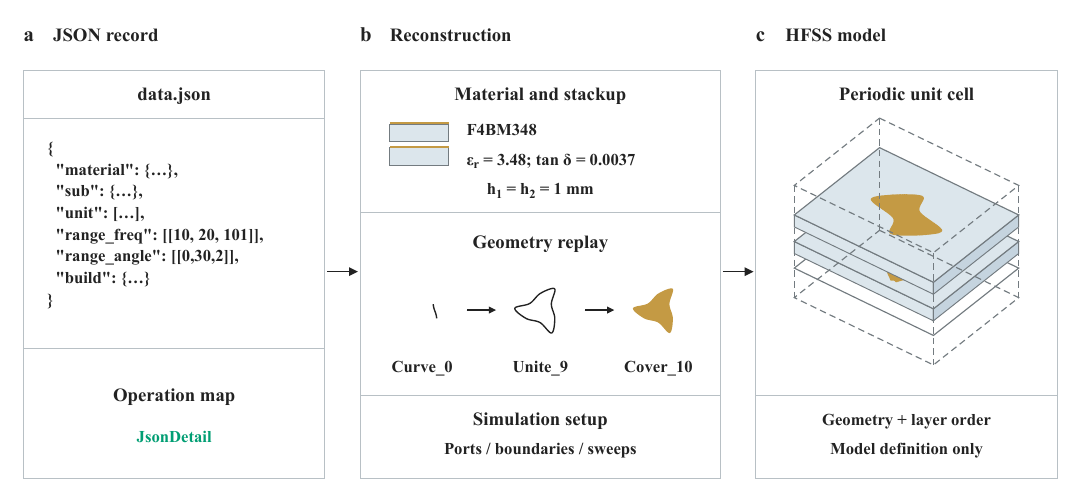}
\caption{Workflow for reconstructing an FSS unit-cell model from record-level JSON metadata. A compact excerpt from record 1786866648.2374983 identifies its material, substrate, unit and sweep metadata together with the ordered build log (a). These inputs support material/stackup restoration, ordered geometry-operation replay and configuration of the reconstructed model through the released code (b). The geometric inset illustrates the recorded first-layer sequence Curve\_0--Curve\_8, Unite\_9 and Cover\_10; the remaining recorded operations complete the other layers. The endpoint is the unit-cell model definition (c). The full operation-to-function mapping is documented in JsonDetail. This historical record contains neither an explicit schema-version field nor a boundary-metadata block; boundary handling therefore depends on the reconstruction implementation and its legacy branch. The stored angular descriptor [0, 30, 2] is reproduced verbatim, whereas the supplied raw response file contains only 0\ensuremath{^{\circ}} and 30\ensuremath{^{\circ}} conditions. The diagram does not establish that a present-day interpretation of that descriptor reproduces the original sweep. No new HFSS reconstruction or re-simulation was performed for this figure, and the illustration makes no claim of spectral agreement.}
\label{fig:4}
\end{figure}

As illustrated in Fig.~\ref{fig:4} , reconstruction proceeds by interpreting the recorded operations in their original order.
The procedure first restores project variables and coordinate-system settings, then creates the recorded rectangles, equation-based curves, and polylines.
Where required, line entities are converted into surfaces, after which geometric transformations, rotational duplication, Boolean union, and object renaming are applied to obtain the recorded structure.
The mapping between JSON fields, build-operation entries, and their corresponding reconstruction functions is documented in JsonDetail in the released code repository.

\section{Data records}
\label{sec:data-records}
FSSDataBase is released as a Hugging Face Dataset under the Apache-2.0 license and is available at [\href{https://huggingface.co/datasets/kkking789/FSSDataBase}{kkking789/\allowbreak FSSDataBase}].
The release contains a relational SQLite database (DataBase.db) and a collection of per-sample directories.
The database provides compact, fixed-length representations of response labels and phases, whereas each sample directory retains the original continuous simulation outputs, reconstruction metadata, and visualization files.
Each sample is assigned a timestamp-derived identifier equal to the elapsed seconds since 00:00:00 UTC on 1 January 1970.
The identifier is stored as id in the database, names the corresponding sample directory, and links the database records to the sample-level files.
The sample\_path field records the directory path associated with each sample.

The SQLite database contains a parent table, samples, and two child tables, responses and repsonses\_angle.
The child tables store one response record for each combination of sample identifier, response direction, and incidence angle.
In both child tables, $direct=0$ denotes $S_{11}$, and $direct=1$ denotes $S_{21}$.
The responses table stores integer response labels, while responses\_angle stores the corresponding phase values.

\begin{table}[htbp]
\centering
\caption{SQLite relational schema.}
\label{tab:1}
\small
\renewcommand{\arraystretch}{1.25}
\begin{tabular}{@{}>{\raggedright\arraybackslash}p{\dimexpr 0.205\linewidth-1.23\tabcolsep\relax}>{\raggedright\arraybackslash}p{\dimexpr 0.155\linewidth-0.9299999999999999\tabcolsep\relax}>{\raggedright\arraybackslash}p{\dimexpr 0.23\linewidth-1.3800000000000001\tabcolsep\relax}>{\raggedright\arraybackslash}p{\dimexpr 0.41\linewidth-2.46\tabcolsep\relax}@{}}
\toprule
\textbf{Table} & \textbf{Role} & \textbf{Primary key} & \textbf{Description} \\
\midrule
samples & Parent table & id & Sample-level identifier, path, lateral size, and total substrate height \\
responses & Child table & (id,direct,angle) & Fixed-length integer-coded response vectors \\
responses\_angle & Child table & (id,direct,angle) & Fixed-length floating-point phase vectors \\
\bottomrule
\end{tabular}
\end{table}

\begin{table}[htbp]
\centering
\caption{samples field dictionary.}
\label{tab:2}
\small
\renewcommand{\arraystretch}{1.25}
\begin{tabular}{@{}>{\raggedright\arraybackslash}p{\dimexpr 0.19\linewidth-0.76\tabcolsep\relax}>{\raggedright\arraybackslash}p{\dimexpr 0.17\linewidth-0.68\tabcolsep\relax}>{\raggedright\arraybackslash}p{\dimexpr 0.64\linewidth-2.56\tabcolsep\relax}@{}}
\toprule
\textbf{Field} & \textbf{SQLite type} & \textbf{Description} \\
\midrule
id & TEXT & Unique timestamp-derived sample identifier \\
sample\_path & TEXT & Path of the sample directory \\
size & REAL & Lateral unit-cell size \\
height & REAL & Total dielectric-stack height \\
\bottomrule
\end{tabular}
\end{table}

\begin{table}[htbp]
\centering
\caption{Child-table field dictionary.}
\label{tab:3}
\small
\renewcommand{\arraystretch}{1.25}
\begin{tabular}{@{}>{\raggedright\arraybackslash}p{\dimexpr 0.16\linewidth-0.96\tabcolsep\relax}>{\raggedright\arraybackslash}p{\dimexpr 0.18\linewidth-1.08\tabcolsep\relax}>{\raggedright\arraybackslash}p{\dimexpr 0.22\linewidth-1.32\tabcolsep\relax}>{\raggedright\arraybackslash}p{\dimexpr 0.44\linewidth-2.64\tabcolsep\relax}@{}}
\toprule
\textbf{Field} & \textbf{Responses type} & \textbf{Responses\_angle type} & \textbf{Description} \\
\midrule
id & TEXT & TEXT & Sample identifier \\
direct & INTEGER & INTEGER & Response direction: 0 for $S_{11}$, 1 for $S_{21}$ \\
angle & REAL & REAL & Incidence angle in degrees \\
p000-p300 & INTGER & REAL & Ordered spectral entries on the database storage grid \\
\bottomrule
\end{tabular}
\end{table}

The code resamples records to its configured storage grid before insertion.
Therefore, users requiring the unmodified continuous HFSS response should use \textit{raw\_result.csv}, rather than the database vectors.
Each sample directory contains machine-readable raw and label files, reconstruction metadata, and visualization files.csv files are comma-delimited and written without a header row.

\begin{table}[htbp]
\centering
\caption{Per-sample directory contents.}
\label{tab:4}
\small
\renewcommand{\arraystretch}{1.25}
\begin{tabular}{@{}>{\raggedright\arraybackslash}p{\dimexpr 0.36\linewidth-1.44\tabcolsep\relax}>{\raggedright\arraybackslash}p{\dimexpr 0.28\linewidth-1.12\tabcolsep\relax}>{\raggedright\arraybackslash}p{\dimexpr 0.36\linewidth-1.44\tabcolsep\relax}@{}}
\toprule
\textbf{Filename or pattern} & \textbf{Format} & \textbf{Content} \\
\midrule
\textit{raw\_result.csv} & Headerless comma-delimited CSV & Continuous raw electromagnetic response \\
\textit{label\_result.csv} & Headerless comma-delimited CSV & Threshold-derived response labels and phases \\
\textit{data.json} & UTF-8 JSON & Reconstruction and simulation metadata \\
\textit{\{$f_{start}$\}\textasciitilde{}\allowbreak{}\{$f_{stop}$\}Ghz\_\{component\}.png} & PNG & Response visualization \\
\textit{structure\_\{idx\}.png} & PNG & Structural visualization \\
\textit{mask\_\{idx\}.csv} & Headerless comma-delimited CSV & Binary structural mask \\
\bottomrule
\end{tabular}
\end{table}

\section{Technical Validation}
\label{sec:technical-validation}

\begin{figure}[htbp]
\centering
\includegraphics[width=\linewidth]{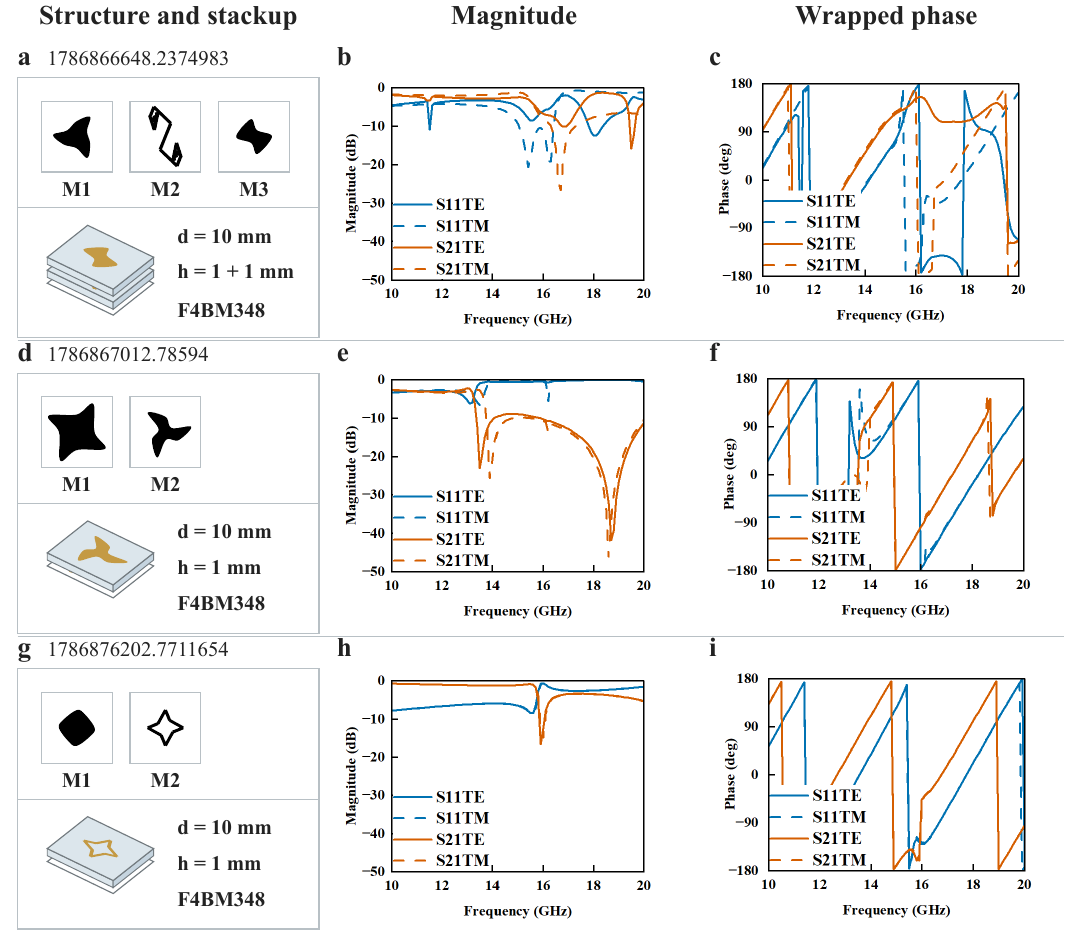}
\caption{Representative FSS structures and their electromagnetic responses.
Three identifiable records illustrate structural and response variation within a bounded inspected subset: 1786866648.2374983 (a--c), 1786867012.78594 (d--f) and 1786876202.7711654 (g--i).
Each row shows the separate metal-layer masks and a schematic layer stack, followed by the corresponding co-polar reflection and transmission magnitudes and wrapped phases at normal incidence.
All curves retain the 101 original frequency samples from 10 to 20 GHz. Blue and vermilion denote S\textsubscript{1}\textsubscript{1} and S\textsubscript{2}\textsubscript{1}, respectively; solid and dashed lines denote TE and TM.
Magnitude is expressed as 20 log\textsubscript{1}\textsubscript{0}|S| in dB and phase as arg(S) in degrees. Black mask regions denote metal. The period is 10 mm and the substrate material is F4BM348 in all three records.
Record A has two 1 mm dielectric layers, whereas B and C have one 1 mm layer, so the examples are not a thickness-controlled performance comparison. The records were chosen for contrasting observed responses from the first 20 inspected directories; they are not a population-wide ranking or a statistical validation of dataset diversity. No curve smoothing, interpolation or phase unwrapping was applied.}
\label{fig:5}
\end{figure}

As illustrated in Fig. ~\ref{fig:5}, the generated dataset contains structurally distinct samples with diverse electromagnetic responses.
We therefore quantified complementary structural and electromagnetic attributes for each valid record to characterize dataset diversity and potential redundancy.
The validation was organized into three complementary analyses: structural diversity, electromagnetic response diversity, and task-orient cost functions for specialized response characteristics.
Together, these analyses characterize the extent of variation and potential redundancy within the dataset, while supporting the identification of records with defined structural or response properties.

\subsection{Structural diversity}
\label{subsec:structural-diversity2}
Structural diversity was quantified from the binary mask $M_{l}\left( i,j\right)\in \left\{ 0,1\right\}$ of each metal layer $l$, where $M_{l}\left( i,j\right)=1$ denotes metal at pixel $\left( i,j\right)$.
For a mask of height $H$ and width $W$, three complementary statistics were calculated.

First, the metal coverage ratio was defined as
\begin{equation}
\label{eq:2}
C_{l}=\frac{1}{HW}\sum_{i=1}^{H} \sum_{j=1}^{W} M_{l}\left( i,j\right).
\end{equation}

This metric describes the fraction of the unit-cell area occupied by metal in layer $l$.

Second, structural eccentricity was measured as the normalized displacement of the metal centroid from the mask central.
The centroid of a non-empty layer is

\begin{equation}
\label{eq:3}
\mu _{l}=\left( \frac{\sum_{i,j} iM_{l}\left( i,j\right)}{\sum_{i,j} M_{l}\left( i,j\right)},\frac{\sum_{i,j} jM_{l}\left( i,j\right)}{\sum_{i,j} M_{l}\left( i,j\right)}\right),
\end{equation}

and the corresponding normalized centroid offset is

\begin{equation}
\label{eq:4}
E_{l}=\frac{||\mu _{l}-c_{0}||_{2}}{\sqrt{{\left( \frac{H-1}{2}\right)}^{2}+{\left( \frac{W-1}{2}\right)}^{2}}},\quad c_{0}=\left( \frac{H-1}{2},\frac{W-1}{2}\right).
\end{equation}

Here, $E_{l}\in \left[ 0,1\right]$; values near zero indicate a metal distribution centered within the unit cell, whereas larger values indicate a more off-central distribution.
Finally, the spatial similarity between two metal layers $l$ and $m$ was evaluated by the intersection-over-union (IoU):

\begin{equation}
\label{eq:5}
IoU_{l,m}=\frac{\sum_{i,j} \left[ M_{l}\left( i,j\right)\cap M_{m}\left( i,j\right)\right]}{\sum_{i,j} \left[ M_{l}\left( i,j\right)\cup M_{m}\left( i,j\right)\right]}.
\end{equation}

For samples containing more than one metal layer, mean pairwise inter-layer IoU was calculated as

\begin{equation}
\label{eq:6}
\overline{IoU}=\frac{2}{L\left( L-1\right)}\sum_{l=1}^{L-1} \sum_{m=l+1}^{L} IoU_{l,m},
\end{equation}

where$ L$ is the number of metal layers.
This statistic is not defined for single-layer samples.
Lower IoU values indicate that metallic patterns occupy more distinct in-plane regions across layers; high IoU values indicate stronger spatial overlap, without by itself implying electromagnetic similarity.

Summary-statistics for the structural-diversity are reported in Table~\ref{tab:5}.
Metal coverage ratio and normalized centroid offset were evaluated at the individual metal-layer level, whereas inter-layer IoU was calculated only for eligible layer pairs within multi-layer samples.
Together, these measures quantify variation in metal occupancy, spatial centering, and cross-layer pattern overlap, respectively.

\begin{table}[htbp]
\centering
\caption{Summary of structural diversity.}
\label{tab:5}
\small
\renewcommand{\arraystretch}{1.25}
\begin{tabular}{@{}>{\raggedright\arraybackslash}p{\dimexpr 0.35\linewidth-2.8\tabcolsep\relax}>{\raggedright\arraybackslash}p{\dimexpr 0.19\linewidth-1.52\tabcolsep\relax}>{\raggedright\arraybackslash}p{\dimexpr 0.18\linewidth-1.44\tabcolsep\relax}>{\raggedright\arraybackslash}p{\dimexpr 0.14\linewidth-1.12\tabcolsep\relax}>{\raggedright\arraybackslash}p{\dimexpr 0.14\linewidth-1.12\tabcolsep\relax}@{}}
\toprule
\textbf{Metric} & \textbf{Records analyzed} & \textbf{Mean$\boldsymbol{ \pm }$ SD} & \textbf{Median} & \textbf{Min-max} \\
\midrule
Metal coverage ratio & 8748 & 0.182$\boldsymbol{ \pm }$0.104 & 0.159 & 0.002-0.611 \\
Centroid offset & 8748 & 9.992e-4$\boldsymbol{ \pm }$1.526e-3 & 6.925e-05 & 0-0.042 \\
Inter-layer IoU & 3644 & 0.245$\boldsymbol{ \pm }$0.2024 & 0.189 & 0-0.994 \\
\bottomrule
\end{tabular}
\end{table}

The distributions reported in Table ~\ref{tab:5} shows that the dataset contains [broad/moderate/limited] variation in the three structural attributes. In particular, [qualitative observation based on the completed table]. These results indicate that the generated records are not dominated by a single metal-coverage, centring, or inter-layer-overlap pattern.

\subsection{Electromagnetic response diversity}
\label{subsec:electromagnetic-response-diversity}

Electromagnetic response diversity was evaluated separately for each response channel $c$ and incidence condition $\theta $, thereby avoiding comparisons between responses obtained under different polarizations, scattering parameters, or angles.
For two samples $a$ and $b$, let $A_{a,c,\theta }\left( f_{k}\right)$ and $\phi _{a,c,\theta }\left( f_{k}\right)$ denote the magnitude in dB and phase in radians, respectively, at the $k$-th aligned frequency point.
The number of shared frequency points is$ K$.

The magnitude-response difference was defined as the root-mean-square difference

\begin{equation}
\label{eq:7}
D_{A}\left( a,b;c,\theta \right)=\sqrt{\frac{1}{K}\sum_{k=1}^{K} {\left[ A_{a,c,\theta }\left( f_{k}\right)-A_{b,c,\theta }\left( f_{k}\right)\right]}^{2}}.
\end{equation}

To account for phase wrapping, the phase-response difference was calculated as

\begin{equation}
\label{eq:8}
D_{\phi }\left( a,b;c,\theta \right)=\sqrt{\frac{1}{K}\sum_{k=1}^{K} {\left[ \operatorname{wrap}_{[-\pi ,\pi )} \left( \phi _{a,c,\theta }\left( f_{k}\right)-\phi _{b,c,\theta }\left( f_{k}\right)\right)\right]}^{2}},
\end{equation}

where $\operatorname{wrap}_{[-\pi ,\pi )} \left( \cdot \right)$ maps a phase difference to the interval $[-\pi ,\pi )$.

The complex S-parameter was reconstructed from magnitude and phase as

\begin{equation}
\label{eq:9}
\tilde{S}_{a,c,\theta }\left( f_{k}\right)={10}^{A_{a,c,\theta }\left( f_{k}\right)/20}\exp  \left[ j\phi _{a,c,\theta }\left( f_{k}\right)\right],
\end{equation}

and the corresponding complex-response distance was defined by

\begin{equation}
\label{eq:10}
D_{S}\left( a,b;c,\theta \right)=\sqrt{\frac{1}{K}\sum_{k=1}^{K} {\left| \tilde{S}_{a,c,\theta }\left( f_{k}\right)-\tilde{S}_{b,c,\theta }\left( f_{k}\right)\right|}^{2}}.
\end{equation}

Finally, the association between magnitude and phase diversity was quantified by the Pearson correlation coefficient across all evaluated sample pairs $\mathcal{P}$:

\begin{equation}
\label{eq:11}
r_{A,\phi }\left( c,\theta \right)=\operatorname{corr} \left( \left\{ D_{A}\left( a,b;c,\theta \right)\right\}_{\left( a,b\right)\in \mathcal{P}},\left\{ D_{\phi }\left( a,b;c,\theta \right)\right\}_{\left( a,b\right)\in \mathcal{P}}\right).
\end{equation}

Summary statistics for the three pairwise distance measures and the magnitude--phase correlation are reported in Table ~\ref{tab:6}.
Magnitude and phase differences are expressed in dB and radians, respectively, whereas $D_{S}$ is dimensionless.
The distance metrics were calculated for $\left[ N_{\mathrm{pair}}\right]$ sample pairs under [response~channels] and [incidence~conditions]; the correlation coefficient was calculated from the same pairwise comparisons.

\begin{table}[htbp]
\centering
\caption{Summary of electromagnetic response diversity.}
\label{tab:6}
\small
\renewcommand{\arraystretch}{1.25}
\begin{tabular}{@{}>{\raggedright\arraybackslash}p{\dimexpr 0.25\linewidth-2.5\tabcolsep\relax}>{\raggedright\arraybackslash}p{\dimexpr 0.2\linewidth-2.0\tabcolsep\relax}>{\raggedright\arraybackslash}p{\dimexpr 0.18\linewidth-1.7999999999999998\tabcolsep\relax}>{\raggedright\arraybackslash}p{\dimexpr 0.14\linewidth-1.4000000000000001\tabcolsep\relax}>{\raggedright\arraybackslash}p{\dimexpr 0.11\linewidth-1.1\tabcolsep\relax}>{\raggedright\arraybackslash}p{\dimexpr 0.12\linewidth-1.2\tabcolsep\relax}@{}}
\toprule
\textbf{Metric} & \textbf{Comparison unit} & \textbf{Records/pairs analyzed} & \textbf{Mean$\boldsymbol{ \pm }$ SD} & \textbf{Median} & \textbf{Min-max} \\
\midrule
Magnitude-response difference & Sample pair & 160000 & 6.875 $\boldsymbol{ \pm }$ 3.029 & 4.752 & 0.002-23.685 \\
Phase-response difference & Sample pair & 160000 & 62.892 $\boldsymbol{ \pm }$ 22.0 & 66.356 & 0.027-155.03 \\
Complex S-parameter distance & Sample pair & 160000 & 0.638 $\boldsymbol{ \pm }$ 0.209 & 0.708 & 0.000-4.396 \\
Magnitude-phase association & Pair-wise-distance set & 160000 & 0.460 $\boldsymbol{ \pm }$ 0.208 & 0.474 & 0.228-0.675 \\
\bottomrule
\end{tabular}
\end{table}

The distributions in Table ~\ref{tab:6}.
indicate [brief qualitative interpretation after calculation, e.g., broad variation in response profiles with a weak/moderate/strong magnitude--phase association].

\subsection{Task-oriented cost functions for specialized response characteristics}
\label{subsec:task-oriented-cost-functions-for-specialized-response-characteristics}

In addition to global structural and electromagnetic-response diversity, task-oriented functions were defined to identify records with selected response characteristics.
These functions are descriptive query metrics rather than application-performance claims.
They were evaluated on aligned frequency samples within a specified scattering channel $c$, polarization $p$, and incidence condition $\theta $.

For any two response traces to be compared, denoted by $x$ and $y$, the normalized magnitude difference and target-phase difference were defined as

\begin{equation}
\label{eq:12}
d_{A}\left( x,y\right)=\sqrt{\frac{1}{K}\sum_{k=1}^{K} {\left[ \frac{A_{x}\left( f_{k}\right)-A_{y}\left( f_{k}\right)}{A_{\mathrm{ref}}}\right]}^{2}},
\end{equation}

\begin{equation}
\label{eq:13}
d_{\phi }\left( x,y;\phi _{0}\right)=\sqrt{\frac{1}{K}\sum_{k=1}^{K} {\left[ \frac{\operatorname{wrap}_{[-\pi ,\pi )} \left( \phi _{x}\left( f_{k}\right)-\phi _{y}\left( f_{k}\right)-\phi _{0}\right)}{\pi }\right]}^{2}},
\end{equation}

where $A_{\mathrm{ref}}=1$dB is a magnitude-scaling constant, $\phi _{0}$ is the target phase difference, and$ K$ is the number of aligned frequency samples.
In the subsequent analyses,$x$ and $y$ correspond to the TE and TM responses, two incidence-angle responses, or the responses of two distinct structures, as appropriate.

\subsubsection{Polarization stability}

For a sample $s$, polarization stability was quantified by the mismatch between the TE and TM responses under the same scattering channel and incidence angle:

\begin{equation}
\label{eq:14}
J_{\mathrm{pol}}\left( s;c,\theta \right)=\frac{1}{2}\left[ d_{A}{\left( S_{s,c,\mathrm{TE},\theta },S_{s,c,\mathrm{TM},\theta }\right)}^{2}+d_{\phi }{\left( S_{s,c,\mathrm{TE},\theta },S_{s,c,\mathrm{TM},\theta };0\right)}^{2}\right].
\end{equation}

Lower values of $J_{\mathrm{pol}}$ indicate more similar TE and TM magnitude-phase responses and, therefore, greater polarization stability.

\subsubsection{Angular selectivity}

Angular selectivity was quantified by the bandwidth over which a sample exhibited a prescribed transmission contrast between two incidence angles, rather than by pointwise magnitude differences alone.
For a selected response channel $c$ and polarization $p$, let $A_{s,c,p,\theta }\left( f\right)$ denote the response magnitude in dB for sample $s$ at frequency $f$ and incidence angle $\theta $.For a directional angular-selection condition, $\theta _{\mathrm{pass}}$ denotes the angle at which transmission is desired and $\theta _{\mathrm{block}}$ denotes the angle at which transmission is suppressed.
Let $L_{s,c,p,\theta }\left( f\right)\in \left\{ -1,1\right\}$ denote the existing discretised response label, where $L=1$ corresponds to a magnitude greater than $-5$ dB and $L=-1$ otherwise.
The binary angular-selection function was defined as

\begin{equation}
\label{eq:15}
q_{\mathrm{ang}}\left( f\right)=\left\{ \begin{matrix}1, & L_{s,c,p,\theta _{\mathrm{pass}}}\left( f\right)=1\ \cap \ L_{s,c,p,\theta _{\mathrm{block}}}\left( f\right)=-1, \\ 0, & \text{otherwise}.\end{matrix}\right.
\end{equation}

Let $\mathcal{F}_{\mathrm{sel}}$ be the set of frequency points for which $q_{\mathrm{ang}}\left( f\right)=1$, and let $\mathcal{I}_{\mathrm{sel}}$ denote the set of contiguous frequency intervals formed from these eligible points.
The angular-selective bandwidth was defined as the width of the longest valid interval:

\begin{equation}
\label{eq:16}
B_{\mathrm{ang}}\left( s;c,p;\theta _{\mathrm{pass}},\theta _{\mathrm{block}}\right)=\max _{I\in \mathcal{I}_{\mathrm{sel}}} \left[ \max _{f\in I} \left( f\right)-\min _{f\in I} \left( f\right)\right].
\end{equation}

If no frequency interval satisfies the selection condition, $B_{\mathrm{ang}}=0$.
The corresponding normalized cost function was defined as

\begin{equation}
\label{eq:17}
J_{\mathrm{ang}}=1-\frac{B_{\mathrm{ang}}}{f_{\max }-f_{\min }},
\end{equation}

where $f_{\min }$ and $f_{\max }$ delimit the analysed frequency range.
Thus, lower values of $J_{\mathrm{ang}}$ correspond to wider continuous angular-selective bandwidths.
In the present dataset, angular selectivity was evaluated using a fixed directional condition: transmission at normal incidence $\left( {0}^{\circ }\right)$ and suppression at oblique incidence $\left( {30}^{\circ }\right)$.
Accordingly, $\theta _{\mathrm{pass}}={0}^{\circ }$ and $\theta _{\mathrm{block}}={30}^{\circ }$ were used throughout the calculation.

\subsubsection{\texorpdfstring{Matched-amplitude structure pairs with a ${\boldsymbol{180}}^{\boldsymbol{\circ }}$ phase difference}{Matched-amplitude structure pairs with a 180 degree phase difference}}

For two different structures $a$ and $b$, a phase-pair mismatch cost was defined to identify response pairs with similar magnitudes and a target phase separation of ${180}^{\circ }$:

\begin{equation}
\label{eq:18}
J_{\pi }\left( a,b;c,p,\theta \right)=\frac{1}{2}\left[ d_{A}{\left( S_{a,c,p,\theta },S_{b,c,p,\theta }\right)}^{2}+d_{\phi }{\left( S_{a,c,p,\theta },S_{b,c,p,\theta };\pi \right)}^{2}\right].
\end{equation}

Lower values of $J_{\pi }$ indicate that the two structures have more closely matched magnitudes and a phase difference closer to ${180}^{\circ }$ over the evaluated frequency range.

Summary statistics for these task-oriented metrics are reported in Table ~\ref{tab:7}.
Polarization stability and angular contrast were calculated over $\left[ N_{\mathrm{condition}}\right]$ valid sample-condition combinations, whereas the ${180}^{\circ }$ phase-pair cost was evaluated over $\left[ N_{\mathrm{pair}}\right]$ eligible structure pairs.
The table contains the numerical results; the surrounding text is limited to their interpretation.

\begin{table}[htbp]
\centering
\caption{Summary of Task-oriented costs.}
\label{tab:7}
\small
\renewcommand{\arraystretch}{1.25}
\begin{tabular}{@{}>{\raggedright\arraybackslash}p{\dimexpr 0.25\linewidth-2.5\tabcolsep\relax}>{\raggedright\arraybackslash}p{\dimexpr 0.23\linewidth-2.3000000000000003\tabcolsep\relax}>{\raggedright\arraybackslash}p{\dimexpr 0.18\linewidth-1.7999999999999998\tabcolsep\relax}>{\raggedright\arraybackslash}p{\dimexpr 0.13\linewidth-1.3\tabcolsep\relax}>{\raggedright\arraybackslash}p{\dimexpr 0.09\linewidth-0.9\tabcolsep\relax}>{\raggedright\arraybackslash}p{\dimexpr 0.12\linewidth-1.2\tabcolsep\relax}@{}}
\toprule
\textbf{Metric} & \textbf{Evaluation unit} & \textbf{Records/pairs analyzed} & \textbf{Mean$\boldsymbol{ \pm }$ SD} & \textbf{Median} & \textbf{Min-max} \\
\midrule
Polarization-stability cost & Sample-condition combination & 10,939 & 0.148 $\boldsymbol{ \pm }$ 0.106 & 0.138 & 0.000–0.595 \\
Angular-selection cost & Sample-condition combination & 10,852 & 0.955 $\boldsymbol{ \pm }$ 0.050 & 0.970 & 0.550-1.000 \\
$\boldsymbol{180^{\circ}}$ phase-pair mismatch cost & Structure pair-condition combination & 800,000 & 1.179 $\boldsymbol{ \pm }$ 0.110 & 1.174 & 0.568-1.819 \\
\bottomrule
\end{tabular}
\end{table}

\section{Usage Notes}
\label{sec:usage-notes}

To facilitate practical reuse of FSSDataBase, we provide a code repository that documents the released generation and reconstruction workflows.
The repository includes the configuration file, instructions for preparing the simulation environment, and utilities for reconstructing and inspecting individual records.
Users are encouraged to inspect the structural representations and corresponding electromagnetic-response files before using the data for model development, evaluation, or design screening.

Individual released samples should be reconstructed from their accompanying \textit{data.json} files using the released reconstruction workflow, rather than by re-running the random generator.
The record-specific JSON metadata preserves the realized geometry, material assignments, layer arrangement, frequency settings, and angular conditions required for this purpose.
Users seeking to reproduce the released generation protocol should retain the versioned code and configuration distributed with the data release without modification.
Re-executing a stochastic generation workflow does not reproduce the original random draws; exact sample-level reconstruction instead relies on the corresponding record metadata.
A licensed, compatible AEDT/HFSS installation and a PyAEDT environment are required for reconstruction and for new simulations.
The documented software configuration should be retained where possible, as equivalence across other software versions is not guaranteed.

The settings.yaml file controls generation of a new, configuration-defined dataset.
Before execution, users should set the dataset and database output locations, the number of attempted samples, the requested number of fixed-grid points stored in each database response vector, and the AEDT version and computational resources available for simulation.
The configuration also specifies the ranges from which unit-cell size and metal trace width are sampled, the available dielectric materials and their electromagnetic properties, and the enabled geometry-generator groups and their associated control parameters.
Layered structures can be generated either by random stack-up construction, for which layer-number ranges, candidate dielectric materials, dielectric thicknesses, and metal-generator candidates can be specified, or by an ordered user-defined arrangement of dielectric and metal layers.
Frequency sweeps are defined as lists of start frequency, stop frequency, and requested sampling points; angular sweeps can be enabled or disabled and are specified by start angle, stop angle, and step size.
Material names referenced by a stack-up must be defined in the material list, and only successfully completed simulation attempts are written as dataset records.
Any modification to these settings defines a new simulation configuration and resulting dataset, and should therefore not be presented as an exact reproduction of the released records.
Implementation-level examples and field-level details are provided in the released documentation.

\section{Code availability}
\label{sec:code-availability}

The source code for generating, reconstructing, and processing FSSDataBase records is available at \url{https://github.com/kkking789/FSSDataBase}.
The version used in this study is release \texttt{V1}, commit \nolinkurl{e6a83d765a46d250af83a0304db30e9554844f5a}.
The workflow was implemented in Python 3.11.
An archived, DOI-versioned copy of this release should additionally be provided where possible.

\section{Data availability}
\label{sec:data-availability}

All data supporting the findings of this study are publicly available through the FSSDataBase repository on Hugging Face: \url{https://huggingface.co/datasets/kkking789/FSSDataBase}.
The dataset is distributed under the Apache-2.0 license and includes sample-level reconstruction metadata, binary structural representations, raw simulated electromagnetic response and associated database files.

\bibliographystyle{unsrtnat}
\bibliography{main}

\end{document}